\documentclass[aps,prresearch,twocolumn,superscriptaddress]{revtex4-2}

\usepackage{amsmath,amssymb,color}

\usepackage{caption}
\usepackage{float}
\usepackage{graphicx}
\usepackage{subfig}

\begin{document}

\title{Stability of $\mathcal{PT}$ and anti-$\mathcal{PT}$-symmetric Hamiltonians with different periodicities} 

\author{Julia Cen}
\email[]{julia.cen@outlook.com}
\affiliation{Theoretical Division and Center for Nonlinear Studies, Los Alamos National Laboratory, Los Alamos, New Mexico NM, USA 87545}

\author{Yogesh N. Joglekar}
\email[]{yojoglek@iupui.edu}
\affiliation{Department of Physics, Indiana University-Purdue University Indianapolis, Indianapolis IN, USA 46202}
	
\author{Avadh Saxena}
\email[]{avadh@lanl.gov}
\affiliation{Theoretical Division and Center for Nonlinear Studies, Los Alamos National Laboratory, Los Alamos, New Mexico NM, USA 87545}
	
\date{\today}
	
\begin{abstract}
Hermitian Hamiltonians with time-periodic coefficients can be analyzed via Floquet theory, and have been extensively used for engineering Floquet Hamiltonians in standard quantum simulators. Generalized to non-Hermitian Hamiltonians, time-periodicity offers avenues to engineer the landscape of Floquet quasi-energies across the complex plane.  We investigate two-level non-Hermitian Hamiltonians with coefficients that have different periodicities using Floquet theory. By analytical and numerical calculations, we obtain their regions of stability, defined by real Floquet quasi-energies, and contours of exceptional point (EP) degeneracies. We extend our analysis to study the phases that accompany these cyclic changes. Our results demonstrate that time-periodic, non-Hermitian Hamiltonians generate a rich landscape of stable and unstable regions.
\end{abstract}
\maketitle


\section{Introduction}
\label{sec:intro}
	
Time-dependent Schr\"{o}dinger equations are in general difficult to solve analytically and usually one has to resort to using perturbation or numerical methods. One of the few analytically solvable time-dependent models is the Rabi model \cite{Rabi1937}, which is a two-level system driven by an oscillatory classical field. Its key feature is that when the field frequency is close to the energy-level gap of the two-level system, the system undergoes complete population inversion at an arbitrarily small field strength. Following the Rabi model, Shirley developed a formalism for periodically driven systems using Floquet theory \cite{Shirley1965, Hanggi1998, Santoro2019}. Over the years, there has been an increasing interest in studying time-periodic driven systems for applications in control and sensing. Examples can be found in the engineering of quantum materials \cite{Oka2019}, topological structures \cite{Kitagawa2010, Lindner2011, Rechtsman2013}, controlling inter-well tunneling of a Bose-Einstein condensate \cite{Lignier2007} and detecting low-frequency magnetic fields \cite{Jiang2021}.
	
During the past quarter of a century, research on non-Hermitian systems incorporating parity-time inversion or $\mathcal{PT}$-symmetries has also seen tremendous growth~\cite{Bender1998,Bender1999}. Such open systems with balanced, spatiotemporally separated gain and loss, are described by effective, non-Hermitian Hamiltonians $H$ with $\mathcal{PT}$ symmetry, $[H,\mathcal{PT}]=0$~\cite{Joglekarepjap}. This antilinear symmetry implies that the Hamiltonian eigenvalues are either real or occur in complex-conjugate pairs, with the $\mathcal{PT}$-symmetry breaking transition occurring at exceptional point (EP) degeneracies where corresponding eigenmodes also coalesce. These remarkable properties have allowed complex extensions of quantum theory \cite{Bender2002, Mostafazadeh2002,  Mostafazadeh2003, Mostafazadeh2010, Znojil2015}. The realization of $\mathcal{PT}$-symmetry as a pair of optical waveguides \cite{Ganainy2007, Klaiman2008, Guo2009, Ruter2010} further fueled the growth of work on the experimental side in a wide range of areas including unidirectional invisibility \cite{Lin2011}, electrical circuits \cite{Schindler2011, Wang2020, Juarez2021}, photonic lattices \cite{Regensburger2012}, mechanics \cite{Bender2013}, optical resonators \cite{Chang2014}, acoustics \cite{Zhu2014}, as well as in minimal quantum systems~\cite{Wu2019,Naghiloo2019,Chen2021,Chen2022,Ding2021}. 
 
A decade ago, new symmetry called anti-$\mathcal{PT}$-symmetry ($\mathcal{APT}$-symmetry) was introduced in non-Hermitian systems and was realized as balanced negative and positive refractive index materials \cite{Ge2013}. A system is said to have $\mathcal{APT}$-symmetry if the $\mathcal{PT}$ operator anti-commutes with the Hamiltonian $H'$, i.e. $\{H', \mathcal{PT}\}=0$. Although, prima facie, the two symmetries appear distinct, it is easy to check that $H$ is $\mathcal{PT}$ symmetric if and only if the Hamiltonian $H'=iH$ has the $\mathcal{APT}$ symmetry. Such Hamiltonians have been realized in flying atoms \cite{Peng2016}, optical waveguides with imaginary couplings \cite{Zhang2019}, optical four-wave mixing in cold atoms \cite{Jiang2019}, Lorentz dynamics \cite{QLi2019},  diffusive systems \cite{YLi2019}, qubits \cite{Cen2022}, nonlinear dimers \cite{Rodrigues2022}, and trapped ions \cite{Bian2022, Ding2022}.
 
When $\mathcal{PT}$-symmetries were introduced for Floquet systems, this brought numerous novel phenomena that fundamentally arise due to the interplay between periodic gain-loss variation and oscillatory, unitary dynamics introduced by a static, Hermitian Hamiltonian. For $\mathcal{PT}$-symmetric Hamiltonians with a single drive---an oscillatory gain-loss term or an oscillatory Hermitian term, but not both--- this led to a rich landscape of $\mathcal{PT}$-symmetric, stable regions and $\mathcal{PT}$-broken unstable regions separated by contours of EPs. This landscape contains $\mathcal{PT}$-broken regions at vanishingly small gain-loss strengths, as well as $\mathcal{PT}$-symmetric regions at arbitrarily large gain-loss strengths, at specific modulation frequencies~\cite{Luo2013,Joglekar2014,Lee2015}. Non-Hermitian Floquet dynamics with $\mathcal{PT}$-symmetries have been investigated in ultracold atoms \cite{Li2019}, nitrogen-vacancy (NV) centers \cite{Liu2021}, superconducting qubits~\cite{Kumar2022, Abbasi2022}, electrical circuits \cite{Chitsazi2017,LeonMontiel2018} and \cite{Kazemi2019,Kazemi2022}, which also includes systems without $\mathcal{PT}$-symmetries. These extensive investigations, however, have been limited to single-frequency drives.  

Here, we investigate $\mathcal{PT}$-symmetric and $\mathcal{APT}$-symmetric Hamiltonians where the Hermitian or gain-loss terms have different periodicities, by using the Floquet formalism. We determine the stable (real Floquet eigenvalues) and unstable (complex conjugate Flqouet eigenvalues) regions numerically by using the  frequency-domain Floquet Hamiltonian $H_F$ and analytically from eigenvalues of the non-unitary, one-period time-evolution operator $G(T)\equiv\exp(-iH_F T)$. Subsequently, we use the cyclic variations of the non-Hermitian Hamiltonians to obtain the bi-orthogonal Berry phase \cite{Berry1984, Zwanziger1990, Sakurai2017} for our models. The plan for the paper is as follows. In Sec.~\ref{sec:floquet} we set out the formalism and establish the notation for both $\mathcal{PT}$-symmetric and $\mathcal{APT}$-symmetric models that we consider. We present the stability phase diagrams for the $\mathcal{PT}$-symmetric case, and discuss their common, salient features, followed by corresponding results for the $\mathcal{APT}$-symmetric models. In Sec.~\ref{sec:berry}, we present the results for the geometric phases acquired during cyclic variations. We conclude the paper with a brief discussion in Sec.~\ref{sec:disc}.


 \section{Floquet formalism}
 \label{sec:floquet}

For a two-level system, a general Hamiltonian is given by $H(t)={\bf A}(t)\cdot\sigma+i{\bf B}(t)\cdot\sigma$ where ${\bf A}(t), {\bf B}(t)$ are real vectors and $\sigma=(X,Y,Z)$ denotes the vector with the Pauli matrices. Since all $2\times 2$ matrices are encoded here, the requisite antilinear symmetry imposes a further constraint of ${\bf A}\cdot{\bf B}=0$. We emphasize that since the Pauli matrices form an orthogonal basis for all $2\times 2$ matrices, arbitrary $\mathcal{PT}$ symmetric and $\mathcal{APT}$-symmetric Hamiltonians can be written in this form. 

For a time-periodic Hamiltonian, the  instantaneous eigenvalues $\lambda_k(t)$ of $H(t)=H(t+T)$ do not determine the dynamics. The long-term dynamics of the system, instead, are determined by the spectrum of the corresponding Floquet Hamiltonian $H_F=H(t)-i\partial_t$ ($\hbar=1$), represented as an infinite-dimensional, non-diagonal matrix in the frequency space. If $H(t)$ contains sinusoidal time-dependence, then the frequency-space Floquet Hamiltonian $H_F$ is tridiagonal~\cite{Joglekar2014,Lee2015}, whereas if $H(t)$ has piecewise constant entries with sharp changes, the matrix $H_F$ has nonzero entries throughout. The spectrum of the Floquet Hamiltonian $H_F(\omega_1,\omega_2)$, comprising Floquet quasienergies $\epsilon_{n\alpha}=\epsilon_\alpha+n\omega$, changes from purely real to complex conjugate pairs as a function of the non-Hermiticity strength $\gamma$ and frequency $\omega\equiv 2\pi/T$ of modulation. An alternate method is to obtain the one-period time-evolution operator $G(T)=\mathbb{T}\exp(-i\int_0^T H(t')dt')$ where $\mathbb{T}$ denotes time-ordered product that takes into account the non-commuting nature of Hamiltonians at different instances of time. The $2\times 2$ non-unitary operator $G(T)$ can also be used to define the Floquet Hamiltonian within the folded-zone scheme~\cite{Joglekar2014,Lee2015} with two Floquet eigenenergies $\epsilon_\alpha$. If one uses piecewise constant Hamiltonian $H(t)$, the operator $G(T)$ can be analytically calculated~\cite{Kumar2022,Harter2020}. 

There are in total three non-trivial cases of $\mathcal{PT}$- and $\mathcal{APT}$-symmetric double-driven models, whereby non-trivial, we mean models with both stable and unstable regions, characterized by real and complex-conjugate Floquet quasi-energies, respectively. In the following paragraphs, we will explore the salient features of each non-trivial case.


\subsection{$\mathcal{PT}$-symmetric Models}
\label{subsec:pt}
Without loss of generality, we consider $\mathcal{PT}$-symmetric models where ${\bf A}$ is in the {\it x-y} plane while ${\bf B}$ is along the $z$-axis. There are many different ways one could implement the double driving, and we start with the following cosineY-cosineZ driven model, 	

\begin{equation}
H\left(t\right)=J X+\gamma\left\{\cos\left(\omega t\right)Y-i \cos\left(\beta \omega t\right) Z\right\}, \label{ptcycztrig}
\end{equation}
where an integer $\beta\geq 1$ denotes that the gain-loss term oscillates faster than the Hermitian term. In the following, we use $J=1$ to set the energy scale, thereby using $\gamma\equiv\gamma/J$ and $\omega\equiv\omega/J$ to denote the dimensionless amplitude and frequency of the Floquet drive. For analytical calculations, we replace the cosine with a step function taking values $\pm 1$, thereby creating a sequence of constant Hamiltonians $H_\ell$ for the $1\leq\ell\leq 4\beta$ equal intervals that span the period $T$. The piecewise constant Hamiltonians are given by 
\begin{eqnarray}
\label{ptcyczpiece}
    H_{\ell} & =& J X+\gamma\left({\bf v}_y\cdot{\bf l}\right) Y-i\gamma\left({\bf v}_z\cdot{\bf l}\right) Z,\\
    {\bf v}_{y} &=& \big(\underbrace{1\cdots 1}_\textrm{$\beta$}\underbrace{-1\cdots -1}_\textrm{$2\beta$}\underbrace{1\cdots 1}_\textrm{$\beta$}\big),\\
    {\bf v}_{z} &=& \big(\underbrace{1,-1,-1,1}_\textrm{$\beta$ copies},\cdots,1,-1,-1,1\big),\\
    {\bf l} &=& (0,\cdots\underbrace{1}_\textrm{$\ell$th entry}\cdots,0).
\end{eqnarray}

	
\begin{figure*}[htbp]
\includegraphics[width=0.95\linewidth]{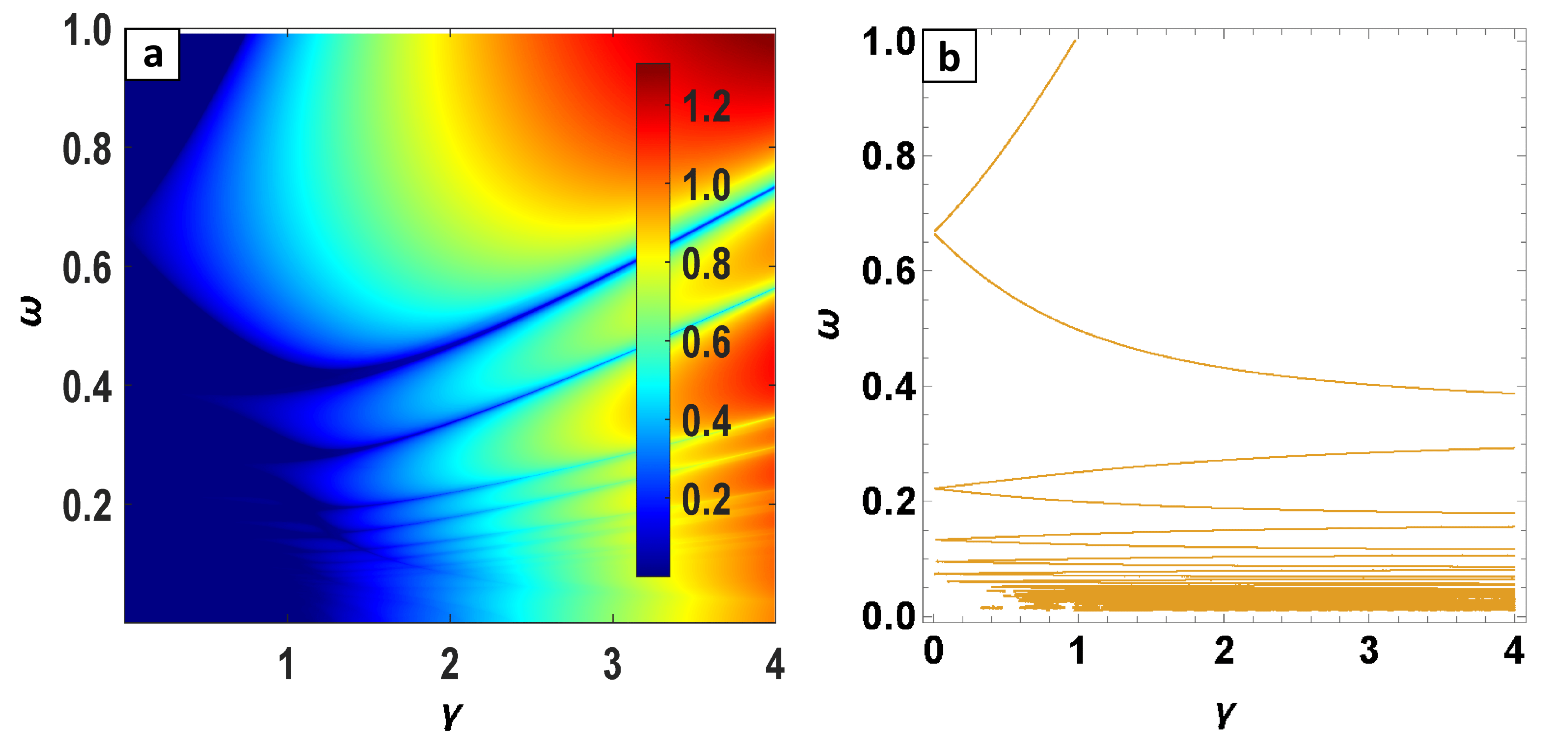}

\caption{Stability of doubly-driven non-Hermitian Hamiltonian with $\beta=3$. (a) Plot of $\max\Im\epsilon_\alpha$ in the parameter space of $\omega$ and $\gamma$ for Hamiltonian (\ref{ptcycztrig}) shows stable deep-blue regions punctuated by unstable regions at $\gamma\ll 1$ at specific frequencies. The resonance at $\omega=2J/\beta$ is clearly seen. (b) EP contours obtained from $G(T)$ for the piecewise constant Hamiltonian (\ref{ptcyczpiece}) also show qualitatively similar features, particularly at small $\gamma$ values. Note that stable regions at large gain-loss strengths $\gamma\gg 1$ are more robust for the piecewise constant Hamiltonian, Eq.(\ref{ptcyczpiece}), than its continuous counterpart, Eq.(\ref{ptcycztrig}).}
\label{fig:f1}
\end{figure*}


Taking Hamiltonian (\ref{ptcycztrig}) and utilizing Floquet theory, we are able to transform our time-dependent problem into an effective time-independent one in which we can find the Floquet quasi-energies, and then obtain their phase diagrams. Constructing the corresponding piecewise version of the driven model (\ref{ptcyczpiece}) allows for analytical calculations to find the phase boundaries or more commonly known as the exceptional lines, dividing unbroken (stable) and broken (unstable) $\mathcal{PT}$-symmetric regions. In the analytical approach, we parametrize the $2\times 2$ Floquet Hamiltonian as $H_{F}=\epsilon_{F}{\bf n}_{F}\cdot\sigma$ where $\epsilon_\alpha=\pm\epsilon_F$ denote the two Floquet quasienergies, ${\bf n}_F$ is a unit vector, and the single-period time evolution operator is given by 
\begin{eqnarray}
G(T)&=&\prod_{\ell=1}^{4\beta}e^{-iH_{\ell}T/4\beta},\\
&=& \cos\left(\epsilon_{F}T\right)\mathbb{I}_{2}-i\sin\left(\epsilon_{F}T\right)\left({\bf n}_{F}\cdot\sigma\right).
\end{eqnarray}
Under this parametrization, the reality of Floquet eigenvalues, which is also the requirement for stability, is equivalent to $\left|\cos\left(4\beta\epsilon_{F}\tau\right)\right|\leq 1$. The EP contours, thus, are given by $\cos\left(4\beta\epsilon_{F}\tau\right)=\pm 1$.

Figure~\ref{fig:f1} shows the representative results of such analysis. The left-hand panel, plotting the maximum imaginary part of Floquet quasi-energies $\max\Im\epsilon_\alpha(\omega,\gamma)$ for $\beta=3$, shows deep-blue stable regions ($\max\Im\epsilon_\alpha=0$) at small gain-loss strengths, punctuated by unstable, $\mathcal{PT}$-symmetry broken regions ($\max\Im\epsilon_\alpha\neq 0$) that occur at vanishingly small $\gamma$ when the modulation frequency $\omega$ takes specific values. We also see thin slivers of stable regions that extend to arbitrarily large gain-loss strengths $\gamma\gg 1$. These results are obtained from an $82\times82$ truncated-matrix $H_F(\omega_1,\omega_2)$ but remain the same when the matrix size is doubled, thereby confirming that the results are valid in the infinite-Floquet-matrix limit. The right-hand panel shows the EP contours that are obtained from the analytical constraint $|\cos(\epsilon_FT)|=1$. It is clear that the structure of the EPs in the $\gamma\ll 1$ region near the primary resonance $\omega=2J/\beta=2/3$ is identical, as is the emergence of many other resonances at low frequencies. On the other hand, the structure of the EP contours at moderate to large values of $\gamma$ is different, although its salient features --- such as the existence of $\mathcal{PT}$-symmetric, stable regions at arbitrarily large values of $\gamma$--- are retained by the analytically tractable model. Similar results are obtained for other integer values of $\beta$, and the phase diagram remains qualitatively similar for rational $\beta$s as well. 

Note that the primary resonance, where unstable regions emerge at vanishingly small gain-loss strength, comes from the anti-Hermitian drive in Eq.(\ref{ptcycztrig}). Our subsequent $\mathcal{APT}$-symmetric models have two anti-Hermitian drives and we will observe two primary resonances, one from each drive. This is a distinguishing feature of systems with different frequency drives. It is also important to point out that when $\beta=1$, the time-dependent part of the Hamiltonian (\ref{ptcycztrig}) is proportional to $(Y-iZ)$. Since the matrix $Y-iZ$ is defective, i.e. it has doubly-degenerate zero eigenvalue with only one eigevector, the eigenvalues of the corresponding Floquet Hamiltonian, with the static $JX$ term, are always real. The non-trivial phase diagram, seen in Fig.~\ref{fig:f1} arises only when $\beta\neq 1$. 


\subsection{$\mathcal{APT}$-symmetric Models}
\label{subsec:apt}

Next, we consider two non-trivial models that are best thought of as $\mathcal{APT}$ symmetric, because they have two anti-Hermitian drives and a constant Hermitian term. We call them cosineX-cosineY and cosineX-sineY models. The sinusoidally varying Hamiltonians for the two cases are given by 
\begin{eqnarray}
H_\textrm{cc}\left(t\right)&=&i\gamma\left\{\cos\left(\omega t\right)X+\cos\left(\beta\omega t\right)Y\right\}+J Z,\label{aptcxcytrig}\\
H_\textrm{cs}\left(t\right)&=&i\gamma\left\{\cos\left(\omega t\right)X+\sin\left(\beta\omega t\right)Y\right\}+J Z.
\label{aptcxsytrig}
\end{eqnarray}
As before, the analysis of the frequency-domain, truncated Floquet Hamiltonian $H_F(\omega_1,\omega_2)$ is supplemented by analytical calculations using piecewise constant Hamiltonians $H_l=i\gamma\left\{({\bf v}_x\cdot{\bf l})X+({\bf v}_y\cdot{\bf l})Y\right\}+JZ$. Here, the corresponding $4\beta$-dimensional vectors ${\bf v}_k$ for the cosineX-cosineY model are given by 
\begin{eqnarray}
    {\bf v}_{x} &=& \big(\underbrace{1\cdots 1}_\textrm{$\beta$}\underbrace{-1\cdots -1}_\textrm{$2\beta$}\underbrace{1\cdots 1}_\textrm{$\beta$}\big),\\
    {\bf v}_{y} &=& \big(\underbrace{1,-1,-1,1}_\textrm{$\beta$ copies},\cdots,1,-1,-1,1\big),
\end{eqnarray}
and those for the cosineX-sineY model are given by 
\begin{eqnarray}
    {\bf v}_{x} &=& \big(\underbrace{1\cdots 1}_\textrm{$\beta$}\underbrace{-1\cdots -1}_\textrm{$2\beta$}\underbrace{1\cdots 1}_\textrm{$\beta$}\big),\\
    {\bf v}_{y} &=& \big(\underbrace{1,1,-1,-1}_\textrm{$\beta$ copies},\cdots,1,1,-1,-1\big).
\end{eqnarray}

\begin{figure}[htbp]
\includegraphics[width=0.95\linewidth]{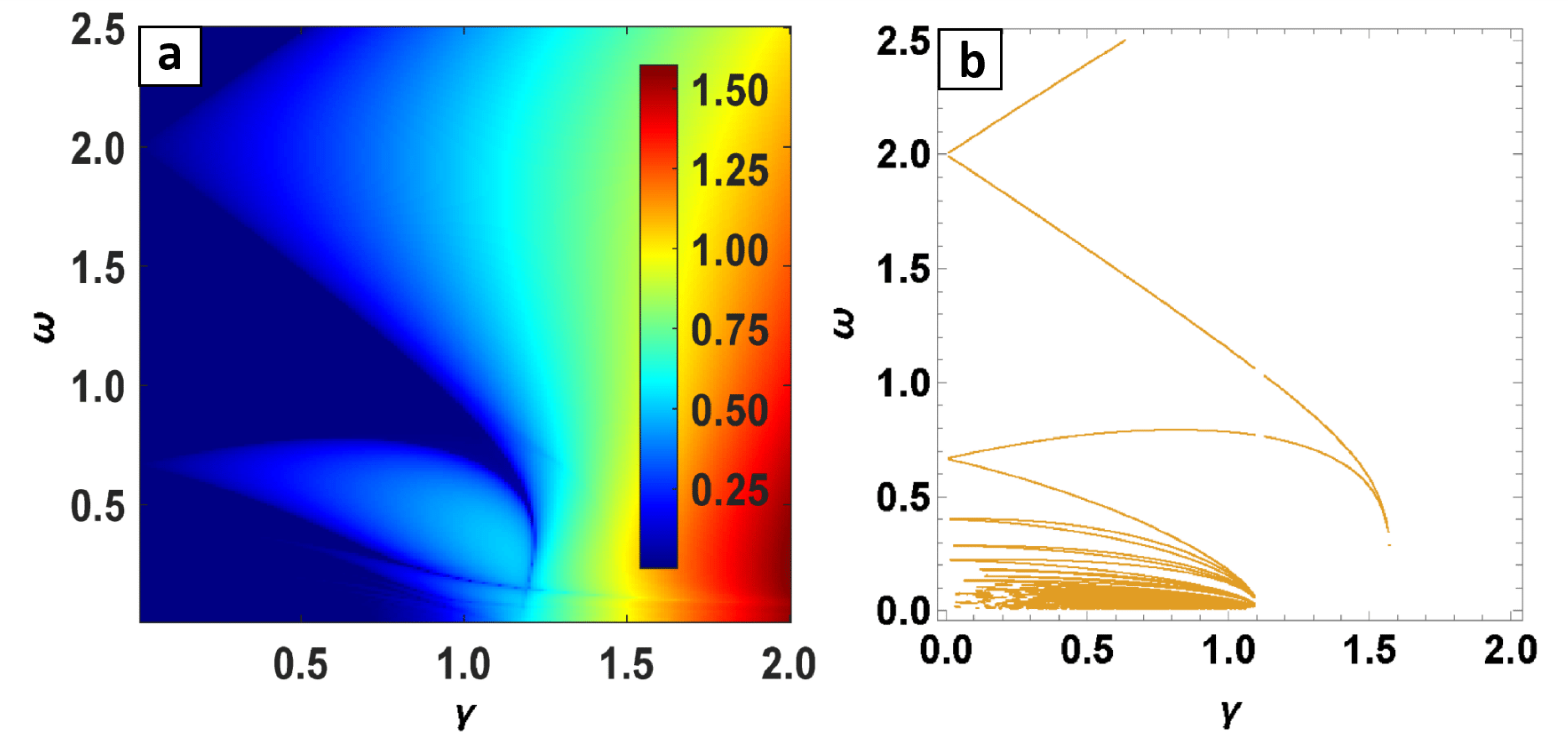}
  \caption{Stability of cosineX-cosineY $\mathcal{APT}$-symmetric Hamiltonian, Eq.(\ref{aptcxcytrig}) with $\beta=3$. (a) $\max\Im\epsilon_\alpha(\gamma,\omega)$ shows triangular unstable regions near resonances $\omega=2,2/\beta$ and their subharmonics. (b) Corresponding EP contours from piecewise-constant Hamiltonian show same qualitative features, including the absence of stable regions at large gain-loss strengths.}
\label{fig:f2}
	\end{figure}

Figure~\ref{fig:f2} shows representative results of such analysis for the Hamiltonian (\ref{aptcxcytrig}) with $\beta=3$. The left-hand panel shows that $\max\Im\epsilon_\alpha$ is zero at small $\gamma\ll 1$ except for the triangular regions of instability ($\max\Im\epsilon_\alpha>0$) that arise near resonances $\omega=2J$, $\omega=2J/\beta$, and its lower subharmonics. We also note, in contrast to the two-drive $\mathcal{PT}$-symmetric case, there are no regions of stability, with real Floquet eigenvalues, at large $\gamma\gg 1$. The right-hand panel in Fig.~\ref{fig:f2} shows the corresponding analytical EP contours, thereby confirming the salient, common features of cosineX-cosineY $\mathcal{APT}$-symmetric Hamiltonian. We note that a perturbation theory approach at small $\gamma$ near the resonances can be used to obtain the slopes and corrections-to-linearity of the triangular unstable regions seen in Fig.~\ref{fig:f2}~\cite{Joglekar2014,Lee2015}. 

\begin{figure}[H]
\includegraphics[width=0.95\linewidth]{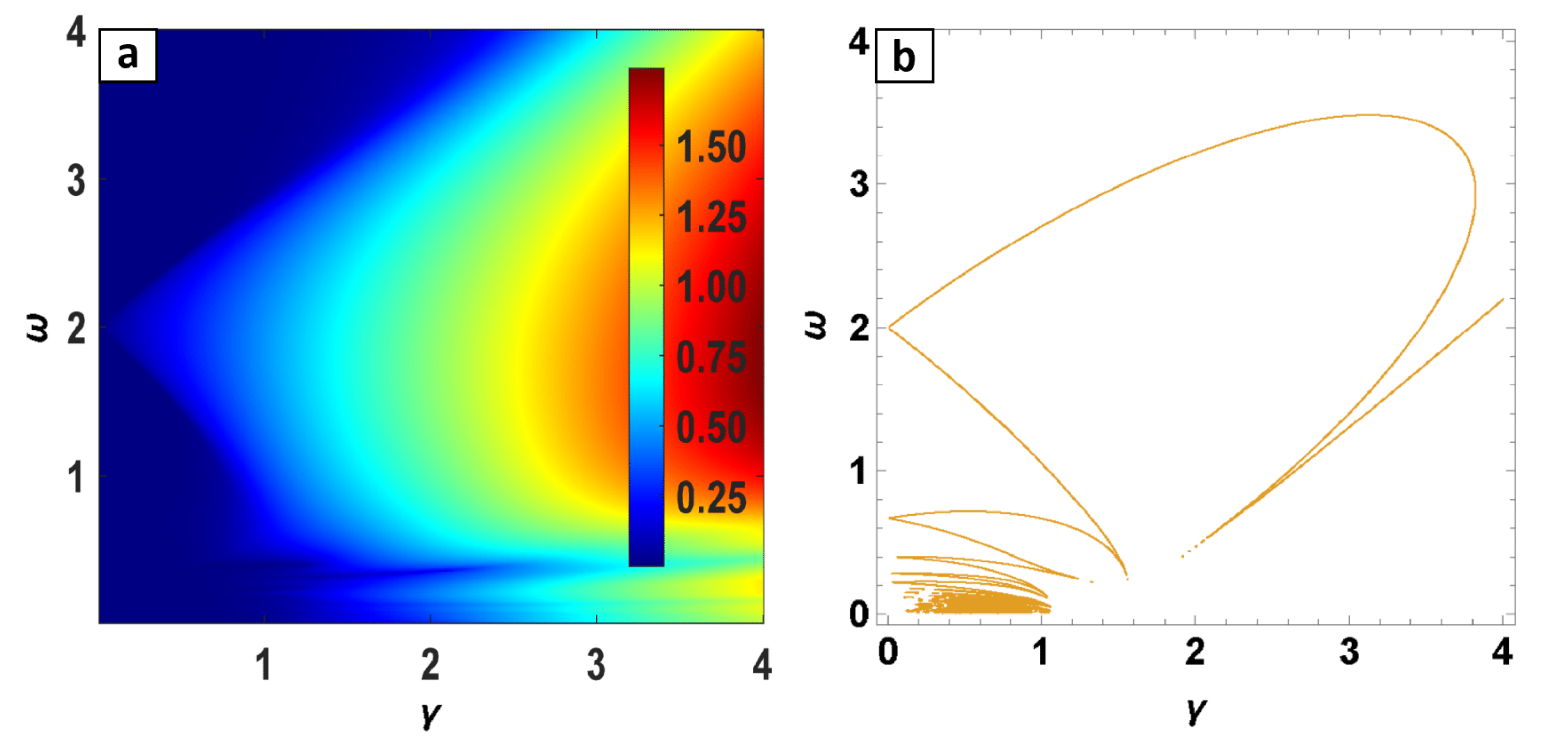}
  \caption{Stability of cosineX-sineY Hamiltonian, Eq.(\ref{aptcxsytrig}) with $\beta=3$. (a) $\max\Im\epsilon_\alpha(\gamma,\omega)$ shows a large triangular unstable region near $\omega=2$ and a small sliver near $\omega=2/\beta$, along with their subharmonics. (b) Corresponding EP contours from piecewise-constant Hamiltonian show same qualitative features. The clustering of EP contours at low frequencies $\omega\ll 1$ to the static threshold $\gamma_\textrm{EP}=1$ is also clear.}
  \label{fig:f3}
	\end{figure}

Results for the cosineX-sineY model, Eq.(\ref{aptcxsytrig}), are shown in Fig.~\ref{fig:f3}. As in the previous cases, the stable regions, indicated by $\max\Im\epsilon_\alpha=0$ are seen at small gain-loss strengths except in the vicinity of primary resonance at $\omega=2J$. The key features of the stability diagram are shared by the EP contours obtained from the piecewise-constant Hamiltonian model. In addition to the large, stable region at high frequencies and moderate gain-loss strengths, the cosineX-sineY model also includes clustering of the low-frequency domain EP contours to the static threshold value of $\gamma_\textrm{PT}=J$. 

\section{Berry Phases under non-Hermitian, cyclic dynamics}
\label{sec:berry}

In Hermitian systems, when the single-drive Rabi problem is generalized to two drives, the system Hamiltonian traces out a closed, two-dimensional loop in the parameter space. Berry phases result from the adiabatic evolution of eigenstates of the Hamiltonian on a closed cycle in the parameter space. The standard Berry-phase expression for Hermitian Hamiltonians is given by
	\begin{equation}
		\theta_{B\alpha}=i \int_{0}^{T} ds \langle \psi_\alpha(s) \vert \partial_s\vert \psi_\alpha(s) \rangle,\label{berry1}
	\end{equation}
where the Hamiltonian $H(s)=H(s+T)$ has eigenstates $|\psi_\alpha(s)\rangle$ with eigenvalues $\epsilon_\alpha(s)$, and Eq.(\ref{berry1}) is valid as long the time $T$ is much longer than inverse of the smallest energy gap $\min|\epsilon_m(s)-\epsilon_n(s)|$ \cite{Berry1984, Zwanziger1990, Sakurai2017}. Due to the unitary evolution generated by $H$, the eigenstates remain normalized, $\partial_s\langle\psi_\alpha(s)|\psi_\alpha(s)\rangle=0$, and therefore the Berry phase defined in Eq.(\ref{berry1}) is a real number. 


\begin{figure*}
\includegraphics[width=0.32\linewidth,height=0.32\linewidth]{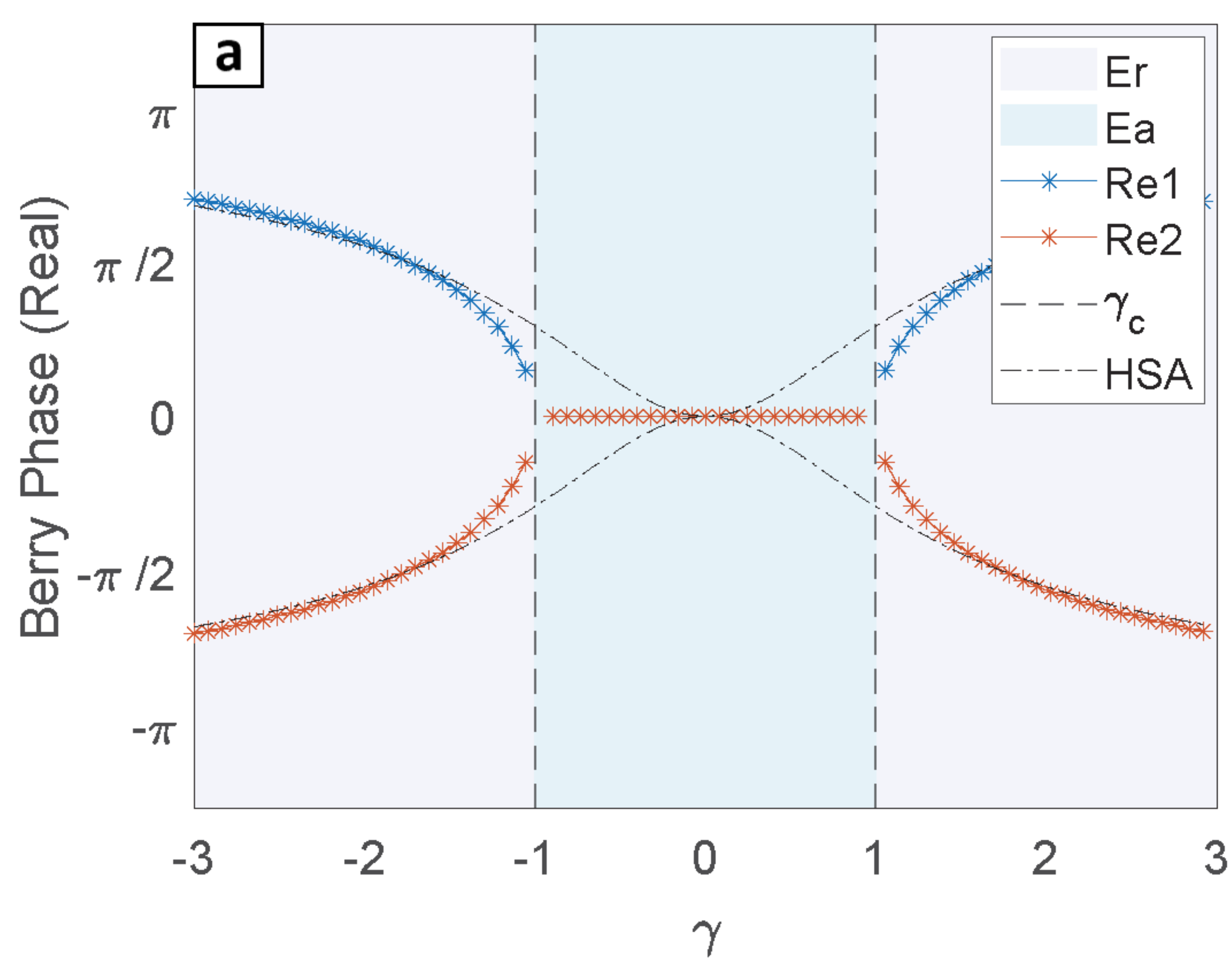}
\includegraphics[width=0.32\linewidth,height=0.32\linewidth]{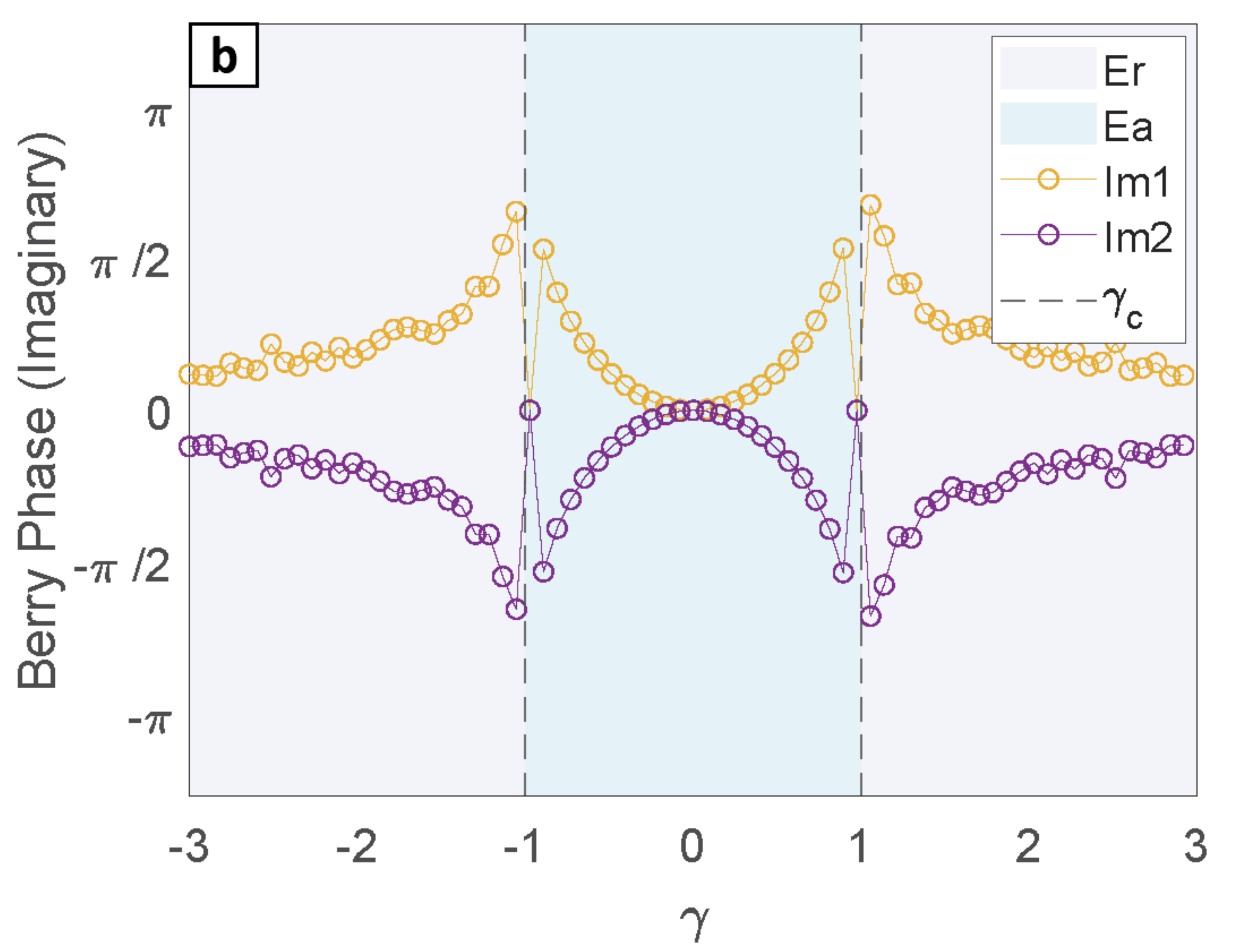}
\includegraphics[width=0.31\linewidth,height=0.32\linewidth]{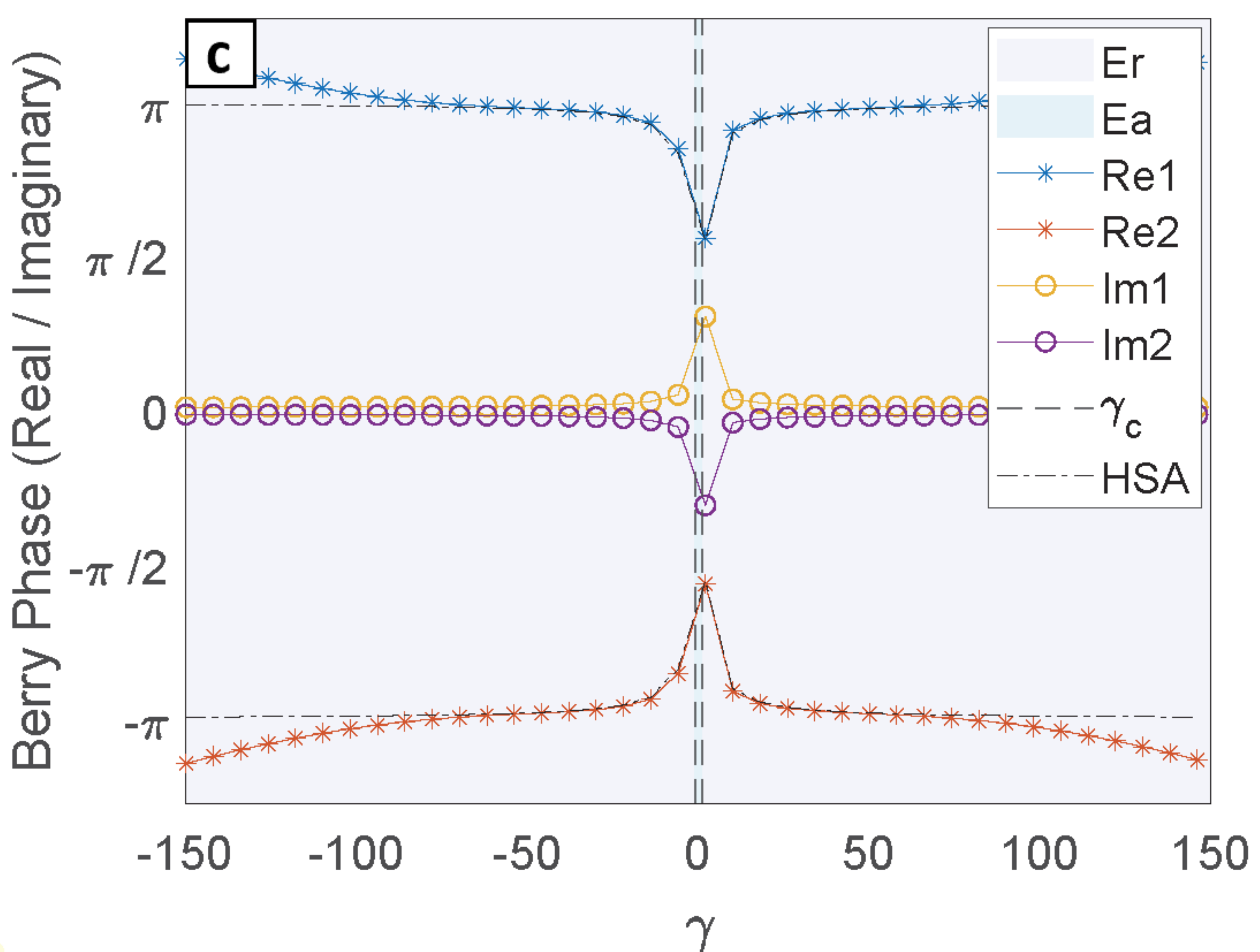}\\
\includegraphics[width=0.32\linewidth,height=0.32\linewidth]{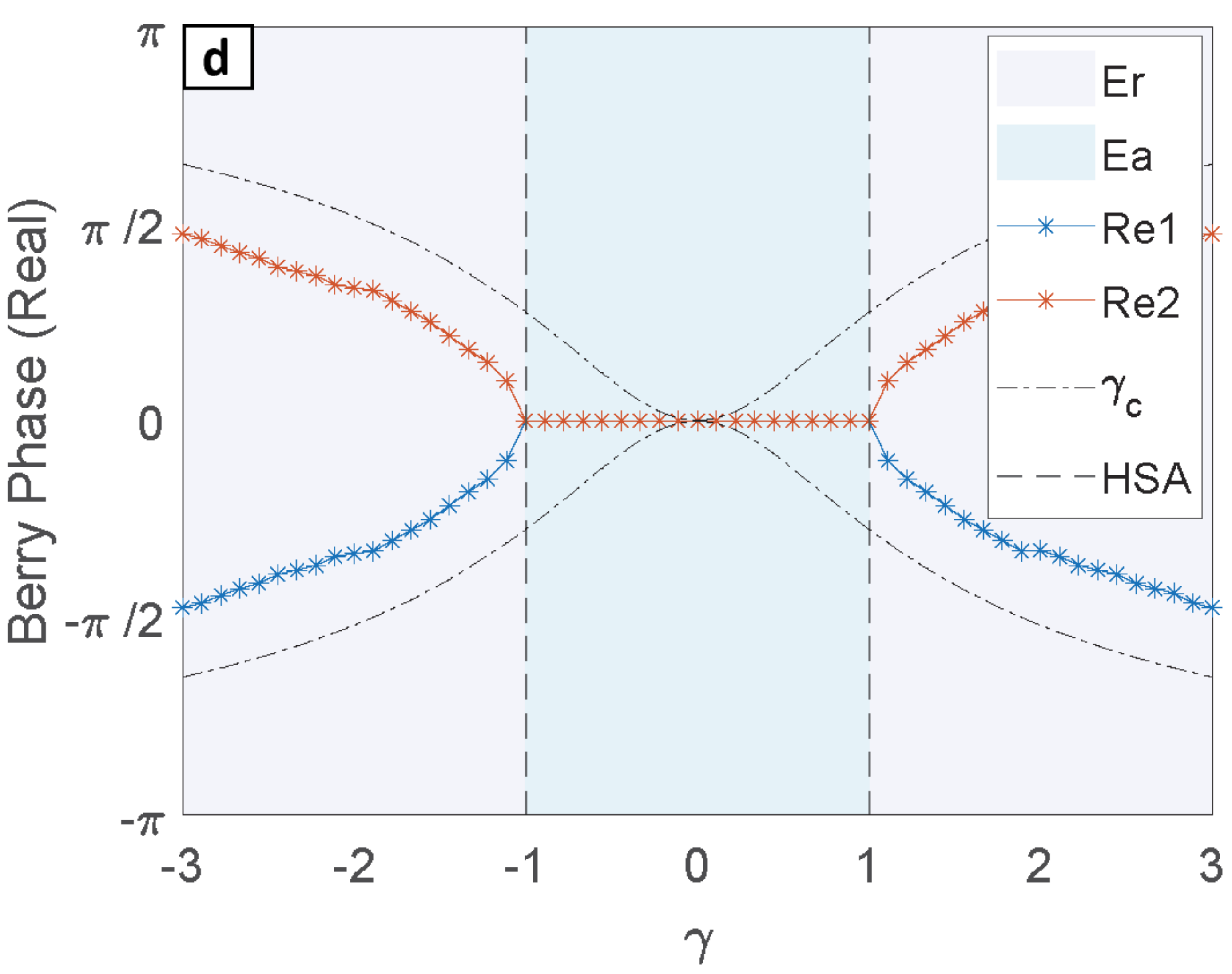}
\includegraphics[width=0.32\linewidth,height=0.32\linewidth]{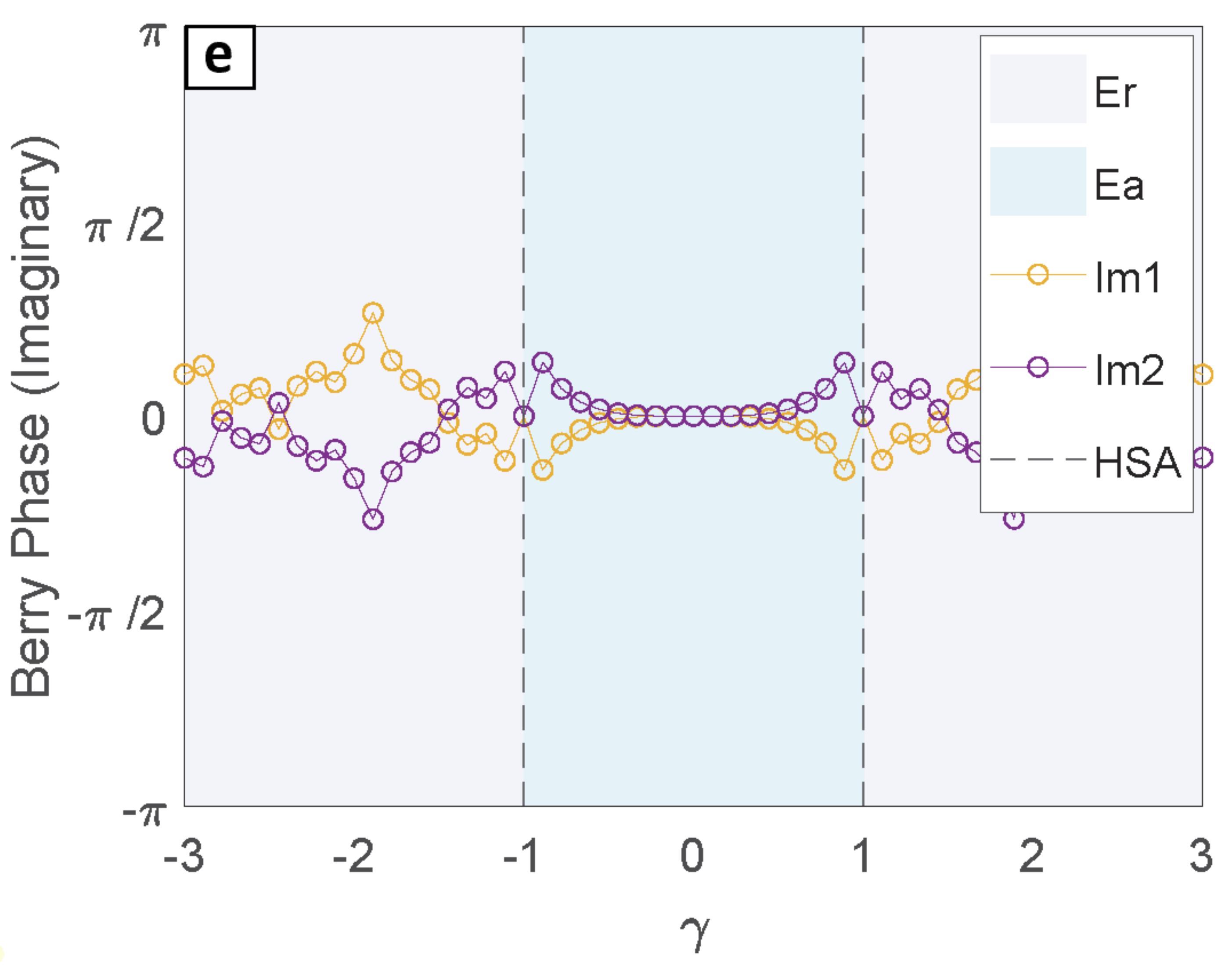}		
\includegraphics[width=0.32\linewidth,height=0.32\linewidth]{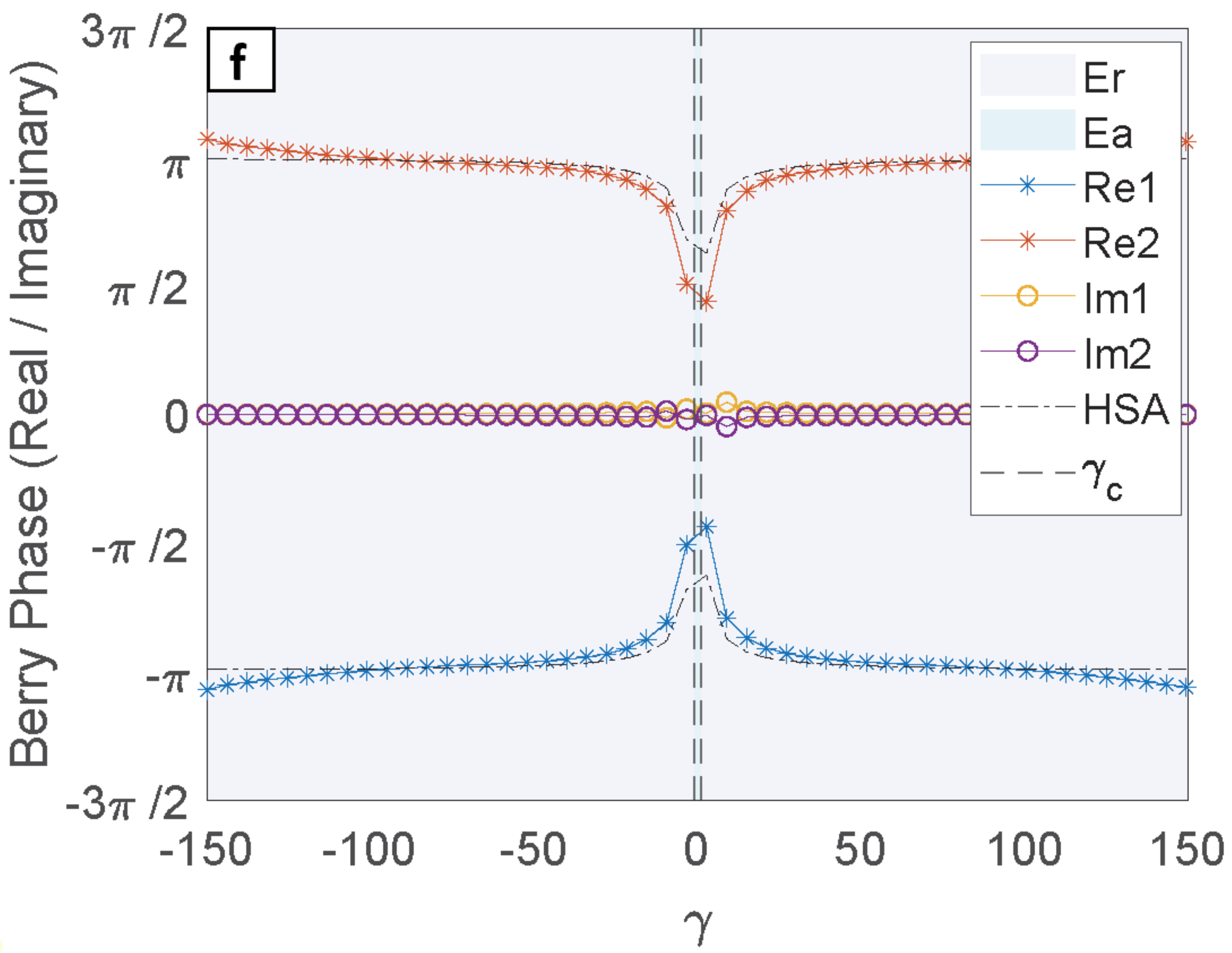}
		
\caption{Real and imaginary parts of the complex Berry phase, Eq.(\ref{berry2}), for the non-Hermitian $\mathcal{PT}$-symmetric system (\ref{ptcysztrig}). (a-c) Top panel shows results for $\beta=1$, i.e. same frequency drives. Blue and red, orange and purple, are the real and imaginary parts of the Berry phases from the two different initial eigenstates of the system. Black dotted-dashed lines denote half the solid angle subtended by the closed curve evolved by the eigenstates. Here, $\gamma_{c}=\pm 1$ denotes the threshold above which the instantaneous eigenvalues become complex for some range of $s\sim T/4,3T/4$. (d-f) Bottom panel shows corresponding results for $\beta=3$.\label{fig:f5}}
\end{figure*}


One interesting feature of doubly-driven non-Hermitian, Floquet models compared with single-driven ones is that we are now able to investigate the analog of  Berry phases under cyclic evolution. Due to left- and right-eigenvectors that are not related by Hermitian conjugation, non-orthogonality of the (left) eigenstates of $H(s)$, the presence of EPs where the minimum energy gap vanishes, and the chiral mode switch that such systems undergo during cyclic evolution~\cite{Liu2021,Abbasi2022}, the notion of Berry phases can be defined in several different ways. We use the biorthogonal approach, where the Berry phase for an eigenstate is defined by
\begin{equation}
		\theta_{B\alpha}=i \int_{0}^{T} ds \langle \psi_{L\alpha}(s) \vert \partial_s\vert \psi_{R\alpha}(s)\rangle,\label{berry2}
	\end{equation}
where $|\psi_{L\alpha}\rangle$ and $|\psi_{R\alpha}\rangle$ denote the biorthonormalized left- and right-eigenstates of the Hamiltonian $H(s)$ with eigenvalue $\epsilon_\alpha$~\cite{Dattolit1990, Sinitsyn2008, Hayward2018, Lu2019}. Since the two are not related by Hermitian conjugation, $\langle\psi_L|\neq |\psi_R\rangle^\dagger$, the quantity defined by Eq.(\ref{berry2}) is a complex number, in general. 

First, we consider the $\mathcal{PT}$-symmetric, cosineY-sineZ model with the Hamiltonian
\begin{equation}
H\left(s\right)=J X+\gamma\left[\cos\left(\frac{2\pi s}{T}\right) Y+ i\sin\left(\frac{2\pi\beta s}{T}\right) Z\right].\label{ptcysztrig}
	\end{equation}
As in the past, we use $J=1$ as the relevant energy scale, and therefore $\gamma\equiv\gamma/J$ denotes the dimensionless strength of the sinusoidal modulation. The top panel in Fig.~\ref{fig:f5} shows the dependence of the complex Berry phase obtained from Eq.(\ref{berry2}) as a function of $\gamma$ at $\beta=1$. Note that the instantaneous eigenvalues of $H(s)$, Eq.(\ref{ptcysztrig}) are given by $\pm\sqrt{J^{2}+\gamma^{2}\cos\left(4\pi s/T\right)}$, and are therefore purely real when $|\gamma|/J\leq 1$. In contrast, for $|\gamma|>J$, the instantaneous eigenvalues become complex-conjugate pairs near $s=T/4,3T/4$. The bottom panel in Fig.~\ref{fig:f5} shows the corresponding results for the complex Berry phase when $\beta=3$. We see that although the behavior at large $|\gamma/J|\gg 1$ is the same in both double-drive cases, details at moderate values of $|\gamma/J|\sim O(1)$ are different. 

\begin{figure}
\includegraphics[width=0.48\linewidth,height=0.48\linewidth]{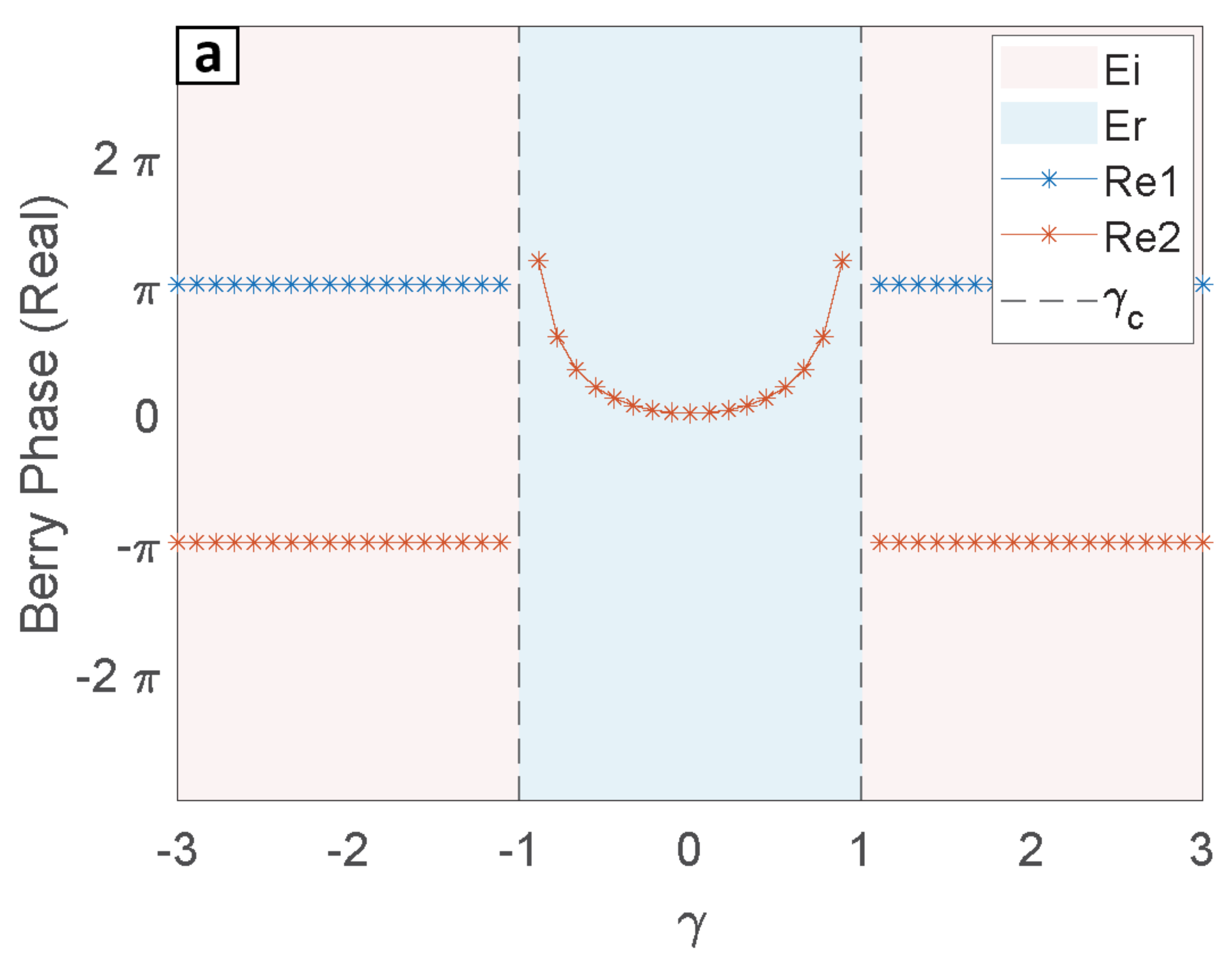}
\includegraphics[width=0.47\linewidth,height=0.47\linewidth]{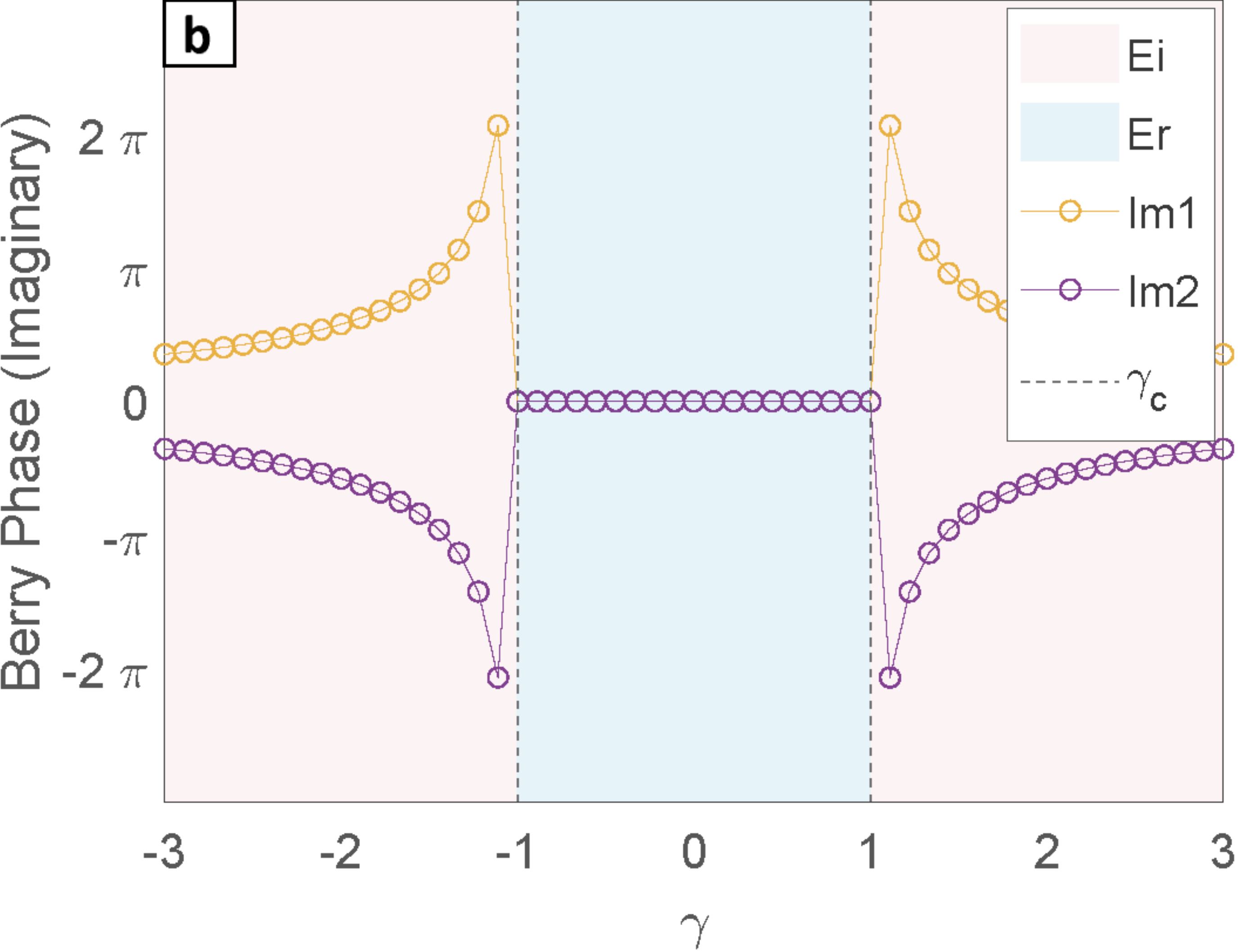}\\
\includegraphics[width=0.48\linewidth,height=0.48\linewidth]{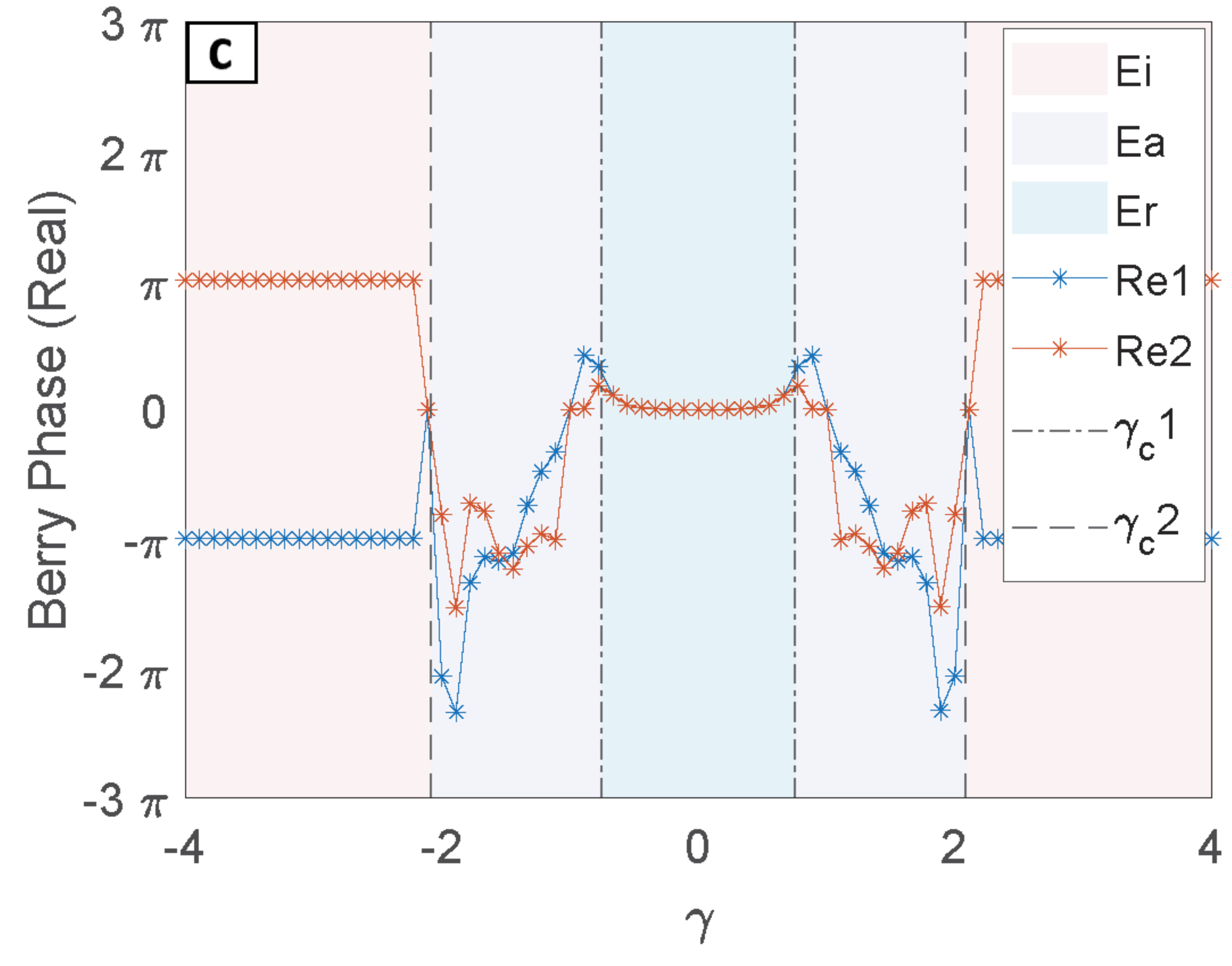}
\includegraphics[width=0.48\linewidth,height=0.48\linewidth]{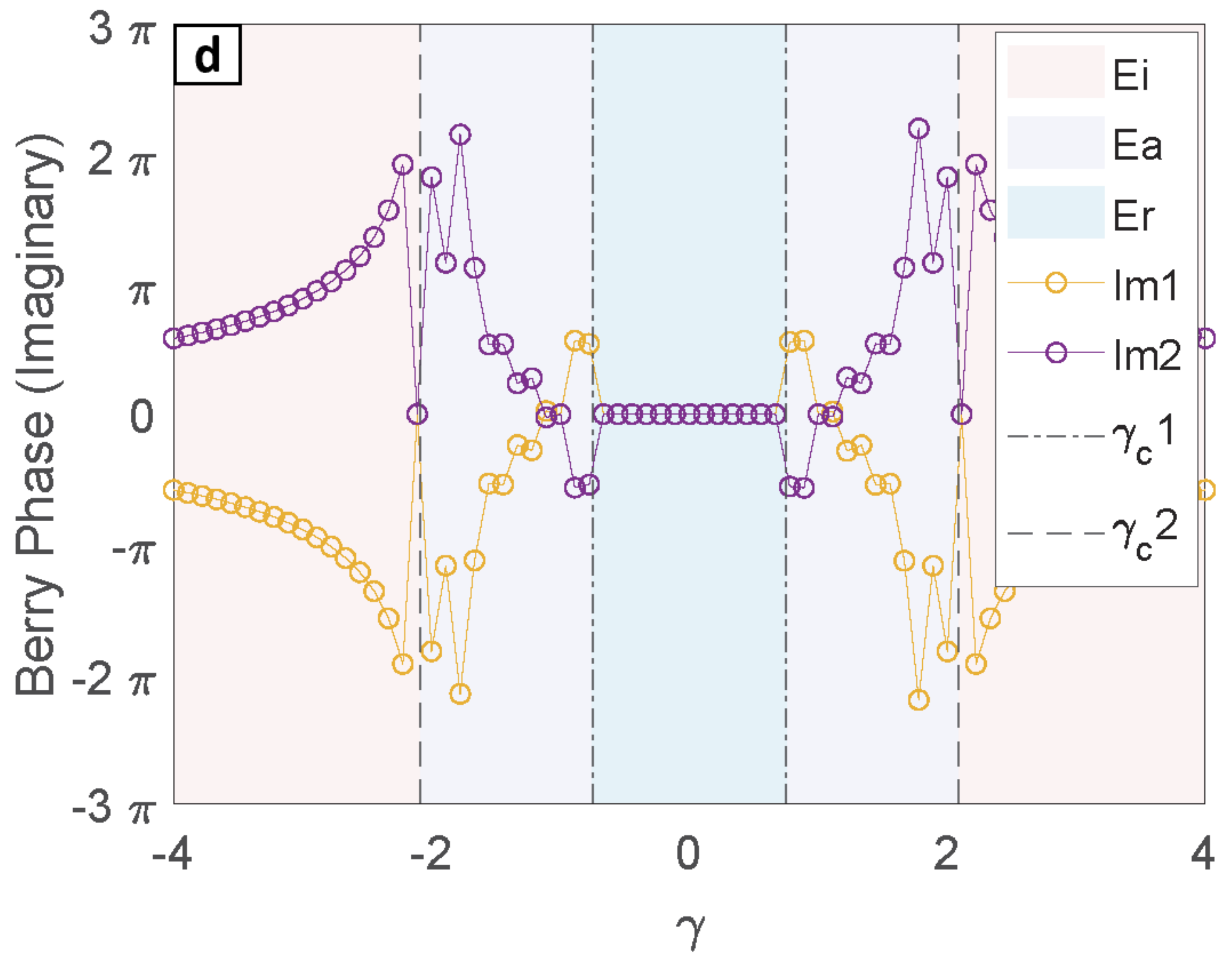}
		
\caption{Real and imaginary parts of the complex Berry phases $\theta_{B\alpha}$ for the $\mathcal{APT}$-symmetric Hamiltonian (\ref{aptcxsy}). (a,b) Top panel shows results for $\beta=1$, i.e. same frequency drives. Blue and red, orange and purple, are the real and imaginary parts of the Berry phases from the two different initial eigenstates of the system. (c,d) Bottom panel shows results for $\beta=3$. Here, $\gamma_{c}$, $\gamma_{c}1$, $\gamma_{c}2$ denote the transition of eigenvalue regions between real (blue region), pure imaginary (pink region) and alternating real and pure imaginary (purple region) cases.\label{F6}}
\end{figure}

Lastly, we consider an $\mathcal{APT}$-symmetric model with two anti-Hermitian drives, 
\begin{equation}        
H\left(s\right)=i\gamma\left[\cos\left(\frac{2\pi s}{T}\right) X+\sin\left(\frac{2\pi\beta s}{T}\right) Y\right]+J Z.\label{aptcxsy}
\end{equation}
When the drives are of the same frequency, $\beta=1$, the instantaneous spectrum of Eq.(\ref{aptcxsy}) is real for $|\gamma|/J\leq 1$ while at larger drive strengths, $|\gamma|/J>1$ the eigenvalues are complex conjugates. The top panel in Fig.~\ref{F6} shows the real and imaginary parts of the complex Berry phase across the threshold $\gamma_c=\pm1$ when $\beta=1$. In this case, the anti-Hermitian drives sweep a circle and therefore we find that the real part of the Berry phase is fixed at $\pm\pi$ when $|\gamma|/J>1$. On the other hand, for smaller drive strengths, the imaginary part of Eq.(\ref{berry2}) is zero. The bottom panel in Fig.~\ref{F6} shows corresponding results when $\beta=3$. Once again, we see the common features where $\Re\theta_{B\alpha}=\pm\pi$ when the instantaneous eigenvalues are complex conjugates, whereas $\Im\theta_{B\alpha}=0$ when they are purely real. 


\section{Discussion}
\label{sec:disc}
	
In this paper, we have analyzed the stability of several $\mathcal{PT}$ and $\mathcal{APT}$-symmetric, non-Hermitian Hamiltonians with different periodicities. We determined the stable and unstable regions using frequency-domain Floquet Hamiltonians for sinusoidal variations, and complemented the analysis by analytical determination of the EP contours by using piecewise constant Hamiltonians. 

Compared with single-drive models, we observe richer dynamics and have more control over the regions of broken $\mathcal{PT}$-symmetry. There are in total only three arrangements of drives with non-trivial Floquet quasi-energies. Although the phase diagrams of Floquet quasi-energies obtained have a lot of similarities with single driving, there are some novel properties. A particular property that emerges when we have driving with different frequencies, is the additional observation of the interlacing of wisps or resonances of the two different frequencies of drives. When we introduce a Hermitian drive into an anti-Hermitian driven system, wisps from the phase diagram do not merge at finite driving strength, $\gamma$. Furthermore, by studying the Berry phases arising from adiabatic cyclic evolution, we find complex structures with non-Hermitian models. Non-Hermiticity brings about an imaginary part to the Berry phases, and in different regions of the Hamiltonian's instantaneous eigenvalues, the behavior of the complex Berry phase for $\mathcal{PT}$-symmetric and $\mathcal{APT}$-symmetric models is markedly different. 

We can classify our models into two categories. All our models have a static Hermitian term and we complement it either with two anti-Hermitian or one Hermitian plus one anti-Hermitian periodic driving. The latter is experimentally more easily implemented, particularly in the semiclassical or quantum platforms where both gain-loss Z-drive and the Hermitian Y-drive are tunable~\cite{Li2019,Wu2019,Liu2021,Naghiloo2019,Abbasi2022,Ding2021}. The former, corresponding to gain-loss Y and Z drives, is challenging: the first one corresponds to Hatano-Nelson model \cite{Hatano1996} while the second one corresponds to the ``standard" gain-loss dimer case. However, non-Hermitian Floquet engineering offers a way to implement both Y and Z anti-Hermitian drives~\cite{Kumar2022}. 

A straightforward extension of this work entails replacing the Pauli operators by higher-dimensional representations of SU(2)~\cite{pr2019,prr2020}. In the static and single-drive Floquet problems, this replacement does not qualitatively change the Floquet results, except that the degree of EPs --- the number of eigenmodes that coalesce together at the degeneracy --- is given by the dimensionality of the representation. Other possible extensions include irrational values of $\beta$ where the total Hamiltonian $H(t)$ is only quasi-periodic. 

This work was supported by the U.S. Department of Energy through the LANL/LDRD Program and the Center for Nonlinear Studies. YNJ acknowledges LANL's hospitality and is supported by ONR Grant No. N00014-21-1-2630.


 
\end{document}